\documentstyle[aps,twocolumn,prl]{revtex}
\def\geqap{\,\raise 2pt \hbox{$>\kern-11pt \lower 5pt \hbox{$\sim$}$}\,}
\def\leqap{\,\raise 2pt \hbox{$<\kern-10pt \lower 5pt \hbox{$\sim$}$}\,}
\input BoxedEPS
\SetTexturesEPSFSpecial
\HideDisplacementBoxes
\begin{document}
\draft
\twocolumn[\hsize\textwidth\columnwidth\hsize\csname @twocolumnfalse\endcsname

\title{ESR investigation on the Breather mode and the Spinon-Breather
dynamical crossover in Cu Benzoate}

\author{T. Asano$^{1}$,  H. Nojiri$^{2}$, Y. Inagaki$^{1}$, J. P.
Boucher$^{1,3}$,
T. Sakon$^{2}$, Y. Ajiro$^{1}$, M. Motokawa$^{2}$}
\address{$^1$Department of Physics, Kyushu University, Fukuoka 812-8581, Japan}
\address{$^2$Institute for Materials Research, Tohoku University, Sendai
980-8577, Japan}
\address{$^3$ Laboratoire de Spectrom$\acute{e}$trie Physique,
Universit$\acute{e}$ J. Fourier, BP 87, F-38402 Saint-Martin
d'H$\grave{e}$res Cedex, France}

\date{\today}
\maketitle
\begin{abstract}
  A new elementary-excitation, the so called "breather excitation", is
observed directly by
millimeter-submillimeter wave electron spin resonance (ESR) in the
Heisenberg quantum spin-chain
Cu benzoate, in which a field-induced gap is found recently by specific
heat and neutron scattering measurements.
Distinct anomalies were found in line width and in resonance field around
the "dynamical crossover"
regime between the gap-less spinon-regime and the gapped breather-regime.
When the temperature becomes sufficiently lower than the energy gap, a new
ESR-line with
very narrow line-width is found, which is the manifestation of the breather
excitation.
The non-linear field dependence of the resonance field agrees well with the
theoretical formula of the first breather-excitation proposed by Oshikawa
and Affleck.
The present work establishes experimentally for the first time that a
sine-Gordon model is applicable to explain spin dynamics
in a $S=1/2$ Heisenberg spin chain subjected to staggered field even in high
fields.

\end{abstract}
]

\narrowtext

 A magnetic field has been recognized as a unique handling-parameter to
control the quantum-critical phenomena in various
low-dimensional spin systems.
An example of the drastic change of magnetic excitation in high magnetic
fields has been
found recently in Cu benzoate Cu(C$_6$H$_5$COO)$_2$$\cdot$3H$_2$O.
For a very long time, this compound had been regarded as a good
representative of $S=1/2$ Heisenberg quantum spin chain
(HQSC), with an exchange coupling $J=8.6$ K.\cite{mdate}
More recently, however, intensive measurements performed below 1 K-specific
heat, neutron scattering
and susceptibility \cite{denderB,denderL}-revealed rather unexpected features.
Besides the dynamical incommensurability expected in high fields, an
unexpected energy-gap {\it E$_g$(H)} in the magnetic
excitation spectrum was observed to develop as a function of the applied
magnetic field {\it H}.
Based on a field theoretical approach, a description was
proposed by Oshikawa and Affleck (OA)\cite{Oshikawa,Affleck} and,
subsequently by Essler and Tsvelik.\cite{EsslerT,Essler}
They claimed that these effects were caused by the staggered
fields acting between neighboring spins in a chain.(Note that some aspects
of the field-induced gap has been discussed theoretically by several
authors,\cite{shultz})
Their most striking theoretical proposal for Cu benzoate is that the
particle-like "breather" excitation appears besides solitons,
in the extreme low-temperature limit, where the temperature is smaller than
the gap.

The "breather" is the soliton-antisoliton bound-state and one of the
elementary excitations in a quantum sine-Gordon model.
For conventional $S=1/2$ HQSC, the gap-less spinon excitation develops due
to the short-range correlation when
the temperature {\it T} is much lower than {\it J}.
In case of Cu benzoate subjected to the staggered field, for further
decrease of the temperature, a drastic change in
spin-fluctuation spectrum arises due to the presence of the field-induced
energy gap; i.e. the dynamical crossover takes place
between the spinon-regime and the gapped breather-regime.
In the gapped breather-regime, it is theoretically proposed by a
sine-Gordon model that the excitation spectrum consist of the first
"breather", soliton, anti-soliton and a multi-particle continuum.
Most important, the exact integrability of this model shows that the ESR
intensity is dominated by the
first-breather mode in the extreme low-{\it T} limit. (Note that, in the
Faraday configuration we used in the experiment, a solition cannot be
excited) For these features, a ESR is considered as one of the most unique
and powerful probes to investigate the spin dynamics in Cu benzoate.

The first "anomaly" of the magnetic excitation below 1 K was reported in ESR
measurements more than 20 years
ago.\cite{Oshima}
Although their interpretation for the low temperature ESR as the
antiferromagnetic resonance was not
compatible with the most recent results as mentioned above, their results
are presently reanalyzed in the newly proposed
theoretical context.\cite{OshikawaE}
It should be noted that no N$\acute e$el ordering was found in the specific
heat and the neutron scattering measurements, at least down to
0.1 K. It means clearly that the field-induced gap should be attributed
purely to the one-dimensional and dynamical character of the
system.

The main issue of the present work is the experimental proof of the
existence of "breather" as well as to elucidate its dynamical properties,
here in Cu benzoate. Our present results also open the possibility to
reinterpret the experimental results on this material obtained more than
twenty years ago.\cite{Oshima} As the first point we test the theoretically
proposed mass-formula of the first-breather ESR mode in
wide range of magnetic field {\it H} including the high-field condition such
as {\it g}$\mu _B${\it H}$\sim
${\it 2J}. As second item we examine by ESR, the dynamical spinon-breather
crossover originating from
the field-induced energy gap in a very wide frequency range.

  For this work, single crystals were grown by the diffusion method and
rectangular shaped samples with a typical dimension of
5$\times$4$\times$0.1 mm$^3$ were used for the measurements.
The quality of crystals was checked by {\it X}-ray diffraction and by
magnetic susceptibility
measurements.
The crystal structure belongs to the monoclinic space group {\it I2/c} and
magnetic chains made up of Cu$^{2+}$ ions are
lined up along the {\it c}-axis.\cite{koizumi}
The details for the ESR system were given in references.\cite{Nojiri,Koyama}
All measurements were performed in the Faraday configuration where a
propagation vector of the light is parallel to the external magnetic field.

    Examples of ESR spectra for ${\it H}{\parallel}{\it c}$ are shown in
Fig. 1 at different frequencies.
Drastic changes in spectra, shift of the resonance field and broadening of
the line width, appear when {\it T}$\leq
${\it J}.
The origin of the {\it T}-dependence is the increase of the spinon
correlation-length $\xi _{spinon}$
that develops as $\xi _{spinon}\sim${\it {J/T}}.
The development of $\xi _{spinon}$ is a common feature of a HQSC and causes
the remarkable changes of ESR spectra.
However, the temperature dependences of the shift and the line width depend
on the type of the anisotropic-term in the
spin-Hamiltonian. For Cu benzoate, it is proposed that a staggered field
gives rise to the anomalies of ESR and that
the shift and the line width follow the relations as $(H/T)^3$ and
$(H/T)^2$, respectively.\cite{OshikawaE}
In fact, as shown in Fig. 1, the shift and the line width drastically
increase as the temperature is decreased.

As the temperature is decreased further, a novel crossover takes place.
While the width of the spinon ESR line S increases continuously, the
integrated intensity of this signal decreases gradually.
At the same time, the new ESR line B$_1$ appears in the low field side.
The lines S and the B$_1$ coexist in some temperature regime and, in this
regime, the spectral weight shifts
gradually from the line S to the line B$_1$. At 0.5 K, finally the
absorption intensity of the B$_1$ becomes dominant.
These behaviors of ESR spectra clearly exhibit that the change should be
considered as "crossover"
rather than a phase transition associated with a well defined phase
boundary. To examine the nature of this crossover, we performed a fit
consisting of two Lorentzian for each spectra and evaluated the width of the
ESR line S and that of the line B$_1$ independently
as a function of temperature as shown in Fig. 2.

The solid lines represent the best fit curves for the ESR line S. The
leading term of the fitting
function is
$\alpha(H/T)^2$, the parameter chosen is as $\alpha=0.087 [K^2 Tesla^{-1}]$.
This term expresses the broadening of the line width caused by the
staggered field, which was proposed by OA in the simplest approximation. In
addition to this main term, a small line width as $\beta H$
($\beta=0.007$, dimensionless) is added, which is caused by other
mechanisms. Surprisingly, the data at all frequencies
are satisfactorily reproduced with only two universal parameters in the wide
temperature range, except for the very vicinity of the
crossover regime.
Since the term $\alpha(H/T)^2$ clearly dominates the line widths, we can say
that the functional form predicted by OA for
the spinon-regime is applicable for a very wide field-temperature range.
It should be noted that the existing low frequency data\cite{Okuda} are
consistently reproduced by using the same parameters we are
proposing here (Note that a small {\it T}-linear term in the previous data
is negligible compared to the $\beta H$ term).
The shift of the resonance field also follows the theoretical
proposal as $(H/T)^3$ mentioned before, although some deviation was found
around the crossover regime. For the limited space of the paper, the
results will be discussed separately.

The line width of the ESR line B$_1$ shows a completely different {\it
T}-dependence from that for the line S.
For the line B$_1$, the width very rapidly decreases as {\it T} is lowered
below the crossover regime.
 (Note that the abscissa of Fig. 2 is a logarithmic temperature scale)
This characteristic {\it T}-dependence can be explained as the effect of the
field-induced gap as follows.
For the gapped breather-regime, the magnetic excitation is dominated by the
particle-like breather-excitation and thus, the ESR
line width is caused by the collisions between the thermally excited
particles. Since the number of excited particles exponentially decreases
toward {\it T}=0, the line width is expected to follow the function as $\it
{\delta H}\propto exp (-\it {E_g(H) }/T)$, which is represented by
the dashed lines in Fig. 2.
Here, we use the experimentally determined gap $\it {E_g(H)}$ from the
resonance fields by using the mass-formula of the first-breather mode(see
Fig. 3(c) for the value of $\it {E_g(H)}$).  We also add a residual line
width $\kappa$ as small as 0.04
Tesla, which is the line width at low-{\it T} limit. This term may come from
the scattering between the defect or impurities
and the breather.
For the prefactor, which relates to the number of breather, we use rather
simply the term $\eta {\it H}$.
Since the transition matrix element to excite the breather mode is the
staggered field,  we can expect that the prefactor is approximately
linear in {\it H}.
It is remarkable that the data at different frequencies are fitted by only
two universal parameters $\eta$ and $\kappa$. Note that $\it {E_g(H)}$
is not an adjustable parameter. The most important point in this analysis is
that the experimentally observed decrease of the line width can be
quantitatively explained by using the experimentally obtained value of the
gap. It means that experimental data, the resonance
fields and the line widths, show the overall agreement with the breather pic
ture proposed by OA. Hence we can conclude clearly that the mode
B$_1$ is the manifestation of the breather excitation.

In the inset of Fig. 2, it is instructive to point out that the crossover
regime, represented by a horizontal bar, is located besides
the solid line which represents the curve $\it {k_BT=E_g(H)}$. This finding
strongly indicates that the drastic anomaly of the
{\it $\delta$H} is caused by the field induced gap:i.e. the line width
probes the essential difference of the spin fluctuation
spectrum between the two regimes with and without an energy gap.
Consequently, it is convincing that the novel anomaly of the ESR signal
observed in the present work is the manifestation of the "spinon-breather
crossover". It should be stressed that the such dynamical
crossover is observed here for the first time directly by means of ESR.

Let us proceed to the discussion of the field dependence of the first
breather mode B$_1$ at the lowest temperature {\it T}=0.5 K.
The frequency-field plots for different field orientations are depicted in
Fig. 3(a).
To examine the field dependence clearly, the deviation {\it $\Delta$} from
the linear Zeeman-effect is depicted in Fig. 3(b) as a
function of external field.

According to OA, the frequency-field relation of the first breather mode
observed by ESR is given by
\cite{OshikawaE}

\begin{eqnarray}
{h \nu}={\sqrt {(g\mu _B H)^2+E_g(H)^2}},
\end{eqnarray}

where $\nu$ and {\it g} are the frequency of ESR and the {\it g}-value,
respectively.
It should be noticed that the field dependence of this breather mode is
different from that observed by neutron scattering
at $\it {q=\pi}$. This difference is originated by the fact that the first
breather mode observed by ESR is a uniform mode at
${\it q=0}$.
The equation means that a non-linear frequency-field relation is caused by
the presence of $\it {E_g}$({\it H}) and thus the
deviation from the liner Zeeman-effect is directly related to the magnitude
of the gap.
This deviation gives rise to the non-linear shift of the resonance field for
the low-field side, which is consistent with the experimental results
as shown in Fig. 3(b).

In Fig. 3(b), it is found that {\it $\Delta$} is large for ${\it
H}{\parallel}{\it c}$ and
${\it H}{\parallel}{\it c''}$, and small for ${\it H}{\parallel}{\it a}$ and
${\it H}{\parallel}{\it b}$ in accordance with the
angular dependence of the field-induced gap.\cite{denderL}
This finding shows that, for ${\it H}{\parallel}{\it a}$ and ${\it
H}{\parallel}{\it b}$, the induced gap is not large enough to satisfy the
condition as ${\it k_BT}\ll$  $\it {E_g}$({\it H}) even at 0.5 K.
Accordingly, we analyze the data only for ${\it H}{\parallel}{\it c}$ and
${\it H}{\parallel}{\it c''}$ in the following.

To calculate  $\it {E_g}$({\it H}) from the experimental data of the
frequency-field dependence, we rewrite the eq(1) as

\begin{eqnarray}
{E_g(H)}=\sqrt {{(h \nu)}^2-(g\mu _BH)^2}.
\end{eqnarray}

By putting the experimental values of $\nu $ and {\it H} and ${\it
g_c}$=2.25 or ${\it g_{c''}}$=2.29 into eq(2), we determine the energy gap
$\it {E_g}$({\it H}) as a function of magnetic field without any adjustable
parameters. It should be noted that this process is very
straightforward as long as the condition ${\it k_BT}\ll$  $\it {E_g}$({\it
H}) is satisfied and thus eq(1) is applicable.
The results are depicted in Fig. 3(c) together with the gap for ${\it
H}{\parallel}{\it c''}$ evaluated by the specific heat
measurements.
The gap obtained by the ESR shows a very good agreement with that obtained
by the specific-heat measurements. It is
remarkable that the validity of the eq(1) is clearly confirmed
experimentally by the present work, in the wide field range including a
high field condition as {\it g}$\mu _B${\it H}$\sim ${\it 2J}.

Finally we briefly discuss the unexpected new ESR lines B$_2$ and B$_3$
observed at 0.5 K. As shown in the inset of Fig. 4,
weak but definite ESR lines B$_2$ and B$_3$ appear besides the first
breather B$_1$.
The slope of B$_2$ and B$_3$ modes are 1.42 {\it g} and 1.65 {\it g},
respectively, where {\it g}=2.25 is the slope of the B$_1$ mode.
It should be noted that such large {\it g}-values of B$_2$ and B$_3$ modes
cannot be attributed to the {\it g}-values of possible
impurities.
Since the energies of B$_2$ and B$_3$ modes are higher than that of the
B$_1$ mode, possible candidates for
these signals are the higher breather-excitations or the transition
associated with multiple breather-excitations.
For a trial, we simply assume that the energy to excite {\it n}-breather is
{\it nE$_g$}({\it H}) and put this mass into eq(1).
The results are shown in Fig. 4 for
{\it n}=1$\sim$4 and some of data points are not so far from the curves.
Although no definite interpretation for the origin
of those extra signals is found at present, we hope that our observation of
the new ESR modes stimulates further theoretical investigations for
the elementary excitation in a quantum sine-Gordon model.

To conclude, a well defined breather excitation has been observed by means
of ESR for the first time, in a very wide field range for Cu
benzoate exposed to a staggered field. The field dependence of the energy
gap agrees well with the results of the
previously reported specific-heat measurements. The present results
establish the validity of the mass-formula of the uniform
first-breather mode at ${\it q=0}$ in a wide field range.
We have also observed the dynamical crossover between the spinon-regime and
the gapped breather-regime.
This crossover takes place when the temperature is comparable to the
magnitude of the gap.
The essential difference of spin fluctuation spectrum between these two
regimes are observed very clearly by the characteristic
temperature dependence of ESR line width.

Prof. M. Oshikawa and Prof. S. Miyashita are acknowledged for many useful
discussions.
This work was partly supported by Grant-in-Aid for Scientific Research from
the Ministry of Education, Science, Sports
and Culture of Japan and by The Sumitomo Foundation.
JPB was supported by a Research Fellowship of the Japan Society for the
Promotion of Science (JPSJ).

%
%
\begin{figure}
%
%
\caption{Examples of ESR spectra at (a) 190 GHz and at (b) 428.9 GHz for
${\it H}{\parallel}{\it c}$. The closed circles and the closed
triangles represent spinon ESR line (S) and the first-breather ESR line
(B$_1$), respectively.}
\label{fig1}
\end{figure}
%
%

%
%
\begin{figure}
%
%
\caption{The temperature dependence of the line width {\it $\delta$H} for
the ESR line S (closed circles) and for the ESR line
B$_1$ (closed triangles).
The line width is defined as a full width at the half maximum.
The solid line and the dashed line represent the functions given in the text.
In the inset, horizontal bars represent the crossover regime where the lines
S and B$_1$ coexist. The
solid line is the curve of {\it k$_B$T}= $\it {E_g}$({\it H}), where value
of  $\it {E_g}$({\it H}) is taken from Fig. 3(c). The dotted line
is an eye-guide.}
\label{fig2}
\end{figure}
%
%

%
\begin{figure}
%
%
%
\caption{(a)Frequency-field plot of the main ESR line for different field
orientations. The marks used in the panel are
identical to those used in other two panels. Thin lines are eye-guides.
(b)The deviation $\Delta$ defined as the difference between the resonance
field at 60 K and that at 0.5 K. The positive
value relates to the shift to the lower field side. The {\it c''}-axis is in
the {\it ac}-plane and tilts
21$^{\circ}$ from {\it a}-axis.(for more detail, see fig. 1 of Ref. [9])
(c)The plot of  $\it {E_g}$({\it H}) as a function of external field.
Open rectangles and closed triangles represent the values obtained by the
present work. Open circles show the value determined by the
specific-heat measurements taken from Ref. [3]. Dashed line and solid line
are the theoretical curves of  $\it {E_g}({\it H})\propto
H^{2/3}$, the prefactor for ${\it H}{\parallel}{\it c''}$ is taken from Ref.
[5].}
\label{fig3}
\end{figure}
%

%
%
\begin{figure}
%
%
%
\caption{Frequency-field plot of the first-breather B$_1$ (closed triangles)
together with the new ESR lines B$_2$ (closed rectangles) and
B$_3$ (closed diamonds) for ${\it H}{\parallel}{\it c}$ and at 0.5 K. Thin
lines are the theoretical curves given in the text. Dashed lines are
eye-guides. An example of ESR spectrum taken at 190 GHz is given in the inset.}
\label{fig4}
\end{figure}
%
%

\vfill
\end{document}